\documentclass[aps,prc,twocolumn,superscriptaddress,showpacs,floatfix,nofootinbib]{revtex4-1}
\usepackage{amsmath,graphicx}
\usepackage{color}
\usepackage[dvipsnames]{xcolor}
\definecolor{mygray}{gray}{0.6}
\usepackage{mciteplus}

\def\P{{\mathcal P}}

\def\tR{\tau_R}

\def\p{{\bf p}}
\def\intg{\int\!\!\!}

\def\Eq#1{Eq.~(\ref{#1})}

\def\Fig#1{Fig.~\ref{#1}}

\def\Ref#1{Ref.~\cite{#1}}

\def\be{\begin{equation}}
\def\ee{\end{equation}}
\def\bea{\begin{eqnarray}}
\def\eea{\end{eqnarray}}

\newcommand \beq{\begin{eqnarray}}
\newcommand \eeq{\end{eqnarray}}
\newcommand{\nn}{\nonumber\\ }
\newcommand{\rme}{{\rm e}}

\begin{document}

\title{ Fluid dynamics  of out of equilibrium boost invariant plasmas}

\author{Jean-Paul Blaizot}
\affiliation{
	Institut de Physique Th{\'e}orique, Universit\'e Paris Saclay, 
        CEA, CNRS, 
	F-91191 Gif-sur-Yvette, France} 
\author{Li Yan}
\affiliation{Department of Physics,
McGill University,
3600 rue University
Montr\'eal, QC
Canada H3A 2T8}
\date{\today}

\begin{abstract}
By solving a simple kinetic equation, in the relaxation time approximation, and for a particular set of moments of the distribution function, we establish a set of equations which, on the one hand, capture exactly the dynamics of the kinetic equation, and, on the other hand, coincide  with the  hierarchy of equations of viscous hydrodynamics, to arbitrary order in the viscous corrections. This correspondence sheds light on the underlying mechanism responsible  for the apparent success of hydrodynamics in regimes that are far from local equilibrium. 

\end{abstract}
\maketitle

\subparagraph{Introduction.}

The observation that the evolution of the quark-gluon plasma produced in ultra-relativistic heavy ion collisions is well described by viscous hydrodynamic equations raises a number of interesting questions that are very much debated presently \cite{Florkowski:2017olj}. Traditional understanding of hydrodynamics would imply that the system has reached local equilibrium, and the small viscosity extracted from the analysis of the data is suggestive of short mean free paths. However, works on strongly coupled plasmas, using in particular holography techniques, indicate that viscous hydrodynamics works even when large anisotropies, that  signal departure from local equilibrium, are still present \cite{Heller:2011ju}. At the same time, there is evidence  that hydrodynamics is capable of describing small colliding systems, for which  no clear separation a priori exists between microscopic and macroscopic scales (see e.g. the recent discussion in \cite{Romatschke:2017vte,Florkowski:2017olj} and references therein). 

Recently, it has been argued that part of the success of hydrodynamics could be due to the existence of a stable attractor, to which the solution of the dynamical equations quickly converge before eventually reaching the viscous hydrodynamic regime \cite{Heller:2015dha}. 
This   suggestion has triggered many studies, some of which involve sophisticated mathematical developments \cite{Spalinski:2017mel,Strickland:2017kux,Romatschke:2017acs,Denicol:2017lxn,Behtash:2017wqg}. In this paper, we would like to offer an alternative  perspective on the issue, based on the simple, and physically motivated  observation, that the main features of the dynamics of expanding plasmas are determined by the competition between the expansion itself, which is dictated by the external conditions of the collisions, and the  collisions among the plasma constituents which generically tend to isotropize the particle momentum distribution functions. These two competing effects give rise to two independent fixed points of a suitably defined dynamical quantity. Many  recent results find a natural interpretation in the interplay between these two fixed points. 

As in many  works on this issue, we focus on the paradigmatic example of the  Bjorken flow \cite{Bjorken:1982qr}, and  consider an expanding system of massless particles 
characterized
by a  distribution function $f$ whose time evolution is given by a kinetic equation. Symmetry allows us to reconstruct the full space-time history of the system from the knowledge of what happens in a slice centered around the plane $z=0$ where the collision takes place. The distribution function in that slice depends  solely on the momentum of the particle and the (proper) time $\tau$, i.e., $f=f(\p,\tau)$. Using  a relaxation time approximation for the collision kernel, we can then write the following simple kinetic equation \cite{Baym:1984np} 
\be
\label{eq:boltz0}
\left[\partial_\tau - \frac{p_z}{\tau}\partial_{p_z}\right] f(\p,\tau) = -\frac{f(\p,\tau)-f_{\rm eq}(p/T)}{\tR}.
\ee 
Here $f_{\rm eq}(p/T)$ is a function that depends only on $p=|\p|$ and an effective temperature $T(\tau)$ which  is determined by requiring that the energy density calculated with $f_{\rm eq}(p/T)$ and $f(\p,\tau)$ takes the same value, $\varepsilon\propto T^4$, at all times. 
The kinetic equation (\ref{eq:boltz0}) makes transparent the competition alluded to above, between the expansion and the collisions.  In the absence of the collision term, the expansion, controlled by the term $-p_z/\tau$ in the left hand side,  leads to a flattened distribution, $f(\p,\tau)\to f_0(p_z\tau,\p_\perp)$, where $f_0$ is the initial distribution and $\p_\perp$ is the component of the momentum orthogonal to the $z$-axis.  On the other hand, the collision term in the right hand side drives the distribution towards isotropy, at a rate controlled by the relaxation time $\tau_R$.

\subparagraph{Kinetics in terms of  ${\cal L}$-moments.}

Although Eq.~(\ref{eq:boltz0}) can be easily solved numerically,  more insight can be gained by using an alternative, albeit approximate, approach that eliminates from the description as much of irrelevant information as possible. Thus, in this paper, instead of considering the full distribution $f(\p,\tau)$, we focus on some of its  moments, introduced in  Ref.~\cite{Blaizot:2017lht}:
\be
\label{eq:moment}
{\cal L}_n = \intg\frac{d^3 \p}{(2\pi)^3p^0}|\p|^2P_{2n}(p_z/|\p|)
f(\p,\tau), 
\ee
where $P_{2n}$ is a Legendre polynomial of order $2n$. The 
moments  ${\cal L}_n$ with   $n\ge 1$ describe  the momentum anisotropy of the system. In particular 
${\cal L}_1=\P_L-\P_T$ reflects the 
asymmetry between longitudinal ($P_L)$ and transverse ($P_T$) pressures. 
The moment  ${\cal L}_0$ coincides with the  energy density, ${\cal L}_0=\varepsilon=P_L+2P_T$. 
Observe that the momentum weight of the integration in Eq.~(\ref{eq:moment}) is always $|\p|^2$, instead of being an increasing power of  $|\p|$ as is the case in more standard approaches (see e.g. ~\cite{Bazow:2016oky}). Thus, the ${\cal L}_n$'s contain little  information on the radial shape of the momentum distribution, preventing us for instance to reconstruct from them the full distribution. However, this  radial shape  plays a marginal role in the isotropization of the momentum distribution, which is our main concern here. Note that all the ${\cal L}_n$  have the same dimension.

By using the recursion relations 
among the Legendre polynomials, we can recast \Eq{eq:boltz0}  into the following (infinite) set of coupled
equations 
\begin{align}
\label{eq:eomL}
&&\frac{\partial {\cal L}_n}{\partial \tau} =& -\frac{1}{\tau}\left[a_n{\cal L}_n
+b_n{\cal L}_{n-1}+c_n{\cal L}_{n+1}\right]-\frac{{\cal L}_n}{\tR}\quad (n\ge 1)\nn
&&\frac{\partial {\cal L}_0}{\partial \tau} =& -\frac{1}{\tau}\left[a_0{\cal L}_0
+c_0{\cal L}_{1}\right],
\end{align}
where 
the  coefficients $a_n,b_n,c_n$ are pure numbers
\begin{subequations}
\begin{align}
a_n=&
\frac{2(14 n^2+7n-2)}{(4n-1)(4n+3)}\,,\quad
b_n=\frac{(2n-1)2n(2n+2)}{(4n-1)(4n+1)}\,,\nonumber\\
c_n=&\frac{(1-2n)(2n+1)(2n+2)}{(4n+1)(4n+3)}, 
\end{align}
\end{subequations}
entirely determined by the free streaming part of the kinetic equation. Note  that the collision term does not affect directly the energy density, but only the moments with $n\ge 1$. In fact, if one ignores the expansion, i.e., set $a_n=b_n=c_n=0$,  the moments evolve according to 
\beq
 {\cal L}_0(\tau)={\cal L}_0(0),\quad 
{\cal L}_n(\tau)={\cal L}_n(0)\,\rme^{-\tau/\tau_R}.
\eeq
This solution illustrates the role of the collisions in erasing  the anisotropy of the momentum distribution as the system approaches equilibrium.  
Of course, the expansion prevents the system to ever reach this trivial equilibrium fixed point: instead, the system goes into an hydrodynamical regime, as we shall discuss later.  

\begin{figure}
\begin{center}
\includegraphics[width=0.35\textwidth] {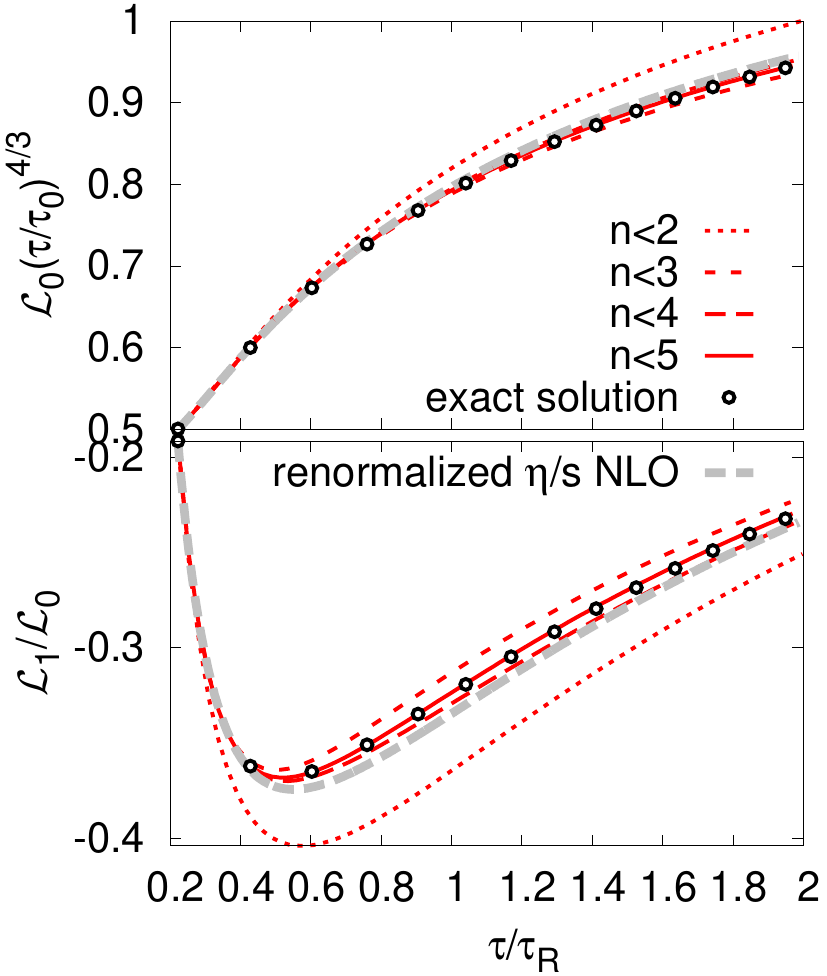}
\caption{(Color online) Comparison of  the ${\cal L}$-moment equations obtained from various truncation of Eqs.~(\ref{eq:eomL}) (lines), with 
those of the numerical solution of the kinetic equation (\ref{eq:boltz0}) (symbols).   \label{fig:compL}
}
\end{center}
\end{figure}

The system of  Eqs.~(\ref{eq:eomL})  lends itself to simple truncations. Thus by ignoring all moments of order higher than $n$, one obtains a finite set of $n+1$ equations that can be easily solved. The accuracy of such a procedure can be judged from \Fig{fig:compL}, where the moments obtained from various truncations are compared with those of the numerical  solution  of Eq.~(\ref{eq:boltz0}) for an initial distribution typical of a heavy ion collision: $f(\tau_0,p_T,p_z)=f_0 \Theta\left(Q_s-\sqrt{\xi^2 p_z^2+ p_T^2}\right)$ with $f_0=0.1$, $\xi=1.5$, corresponding to an initial
momentum anisotropy $\P_L/\P_T\approx 0.5$, and $\tau_0=Q_s^{-1}$ \cite{Blaizot:2017lht}.  
Already the lowest order truncation at $n=1$ captures the qualitative behaviour of the full solution.  Note that the approach to  the exact solution is alternating, which offers an estimate of the truncation error. The energy density approaches smoothly the hydrodynamic regime as $\tau\gtrsim \tau_R$, while the non monotonous behaviour of the ratio ${\cal L}_1/{\cal L}_0$ reflects the competition between expansion and collisional effects that we  now analyze in more detail, starting with the free streaming regime.

\subparagraph{The free streaming fixed point.}

The free streaming regime is described by Eq.~(\ref{eq:eomL}) where one ignores the collision term. It is not hard to see that the resulting equation possesses a stable solution at large time, in which all moments decay as $1/\tau$ and are proportional to each other: ${\cal L}_n(\tau)=A_n{\cal L}_0(\tau)$, where the dimensionless constants $A_n$ characterize the moments of a distribution that is flat in the $p_z$ direction \cite{Blaizot:2017lht}
\beq\label{Ans}
A_n=P_{2n}(0)=(-1)^n\,\frac{(2n-1)!!}{(2n)!!}.
\eeq
Note that $A_1=-1/2$, corresponding to a vanishing longitudinal pressure. 
As for the factor $1/\tau$ it reflects the conservation of the energy in the increasing comoving volume ($\tau \varepsilon(\tau)={\rm cste}$). Defining 
\be
g_n(\tau)=
\tau\partial_\tau\ln {\cal L}_n, 
\ee 
we get from 
\Eq{eq:eomL} 
\be
\label{eq:eomL1}
g_n(\tau)= -a_n - b_n\frac{{\cal L}_{n-1}}{{\cal L}_n} - c_n\frac{{\cal L}_{n+1}}{{\cal L}_n} - (1-\delta_{n0})\frac{\tau}{\tau_R}.
\ee
The solution above corresponds to a fixed point for the $g_n$'s. Dropping the last term, and using the expression (\ref{Ans}) for the ratio of moments, one indeed verifies easily that for all $n$,  $g_n(\tau)=-1$.
If the initial  ratios of moments  are chosen according to Eq.~(\ref{Ans}), the $g_n$'s remain constant in time (all equal to $-1$), whereas for arbitrary initial conditions, they will reach the fixed point at late time. Note that the fixed point  obtained from a truncation at a finite order differs slightly from $-1$: for instance, in the simplest truncation at $n<2$,  $g_0=g_1=-0.92937$ instead of -1, and $A_1\approx -0.6$ instead of $-0.5$.

\subparagraph{The hydrodynamic fixed point.}

We know from our previous study \cite{Blaizot:2017lht} that, at late times, ${\cal L}_n(\tau)$  admits the following expansion, analogous to a gradient expansion\footnote{For Bjorken flow, the gradient expansion coincides with an expansion in powers of $\tau_R/\tau$, which may also be viewed as an expansion in Knudsen number.}
\be
\label{eq:lexp}
{\cal L}_n(\tau)=\frac{1}{\tau^n} \sum_{m=0}^{\infty} \frac{\alpha_n^{(m)}}{\tau^m}\,.
\ee
The coefficients in \Eq{eq:lexp} are nothing but transport
coefficients, except for the first moment, 
equal to the energy density, i.e., $\alpha_0^{(m)}=\varepsilon\delta_{m0}$.  The behavior of $\varepsilon(\tau)$ at large time is obtained from Eq.~(\ref{eq:eomL}), ignoring the contribution of ${\cal L}_1$. Since $a_0=4/3$, this behavior is that of ideal hydrodynamics, $\varepsilon(\tau)\sim \tau^{-4/3}$, and hence $T(\tau)\sim \tau^{-1/3}$.
The leading and sub-leading transport coefficients in Eq.~(\ref{eq:lexp}) can be 
determined analytically. To do so, we return to Eq.~(\ref{eq:eomL1}) and note that a cancellation of the relaxation term has to occur in order to eliminate the exponential decaying contributions to the moments. This cancellation determines the leading order coefficient, viz.   
$\alpha_n^{(0)}=(-\tR)^n \varepsilon \prod_{i=1}^n b_i $. In particular, $\alpha_1^{(0)}=-b_1\tau_R\varepsilon=-2\eta$, with $\eta$ the shear viscosity.  In a conformal invariant setting \cite{Baier:2007ix},  we allow $\tR$ to depend on
 the temperature, with  $\tau_R T(\tau)$  kept constant\footnote{The constant is given by $\tau_R T(\tau)=5\eta/s$, with the entropy density given by   $s=4\varepsilon/(3T)$.}.  Then, one gets   $\alpha_n^{(0)}\sim \tau^{-(4-n)/3}$ which implies that in leading order,  ${\cal L}_n(\tau)\sim \tau^{-(4+2n)/3}$. This defines the hydrodynamic fixed point,  $g_n(\tau)=-(4+2n)/3$.\footnote{In the conformal invariant setting, this result could also be obtained from a simple dimensional analysis. For a  time-independent relaxation time,  the hydrodynamic fixed point is  instead $g_n(\tau)=-(4+3n)/3$.}
The  sub-leading coefficients in Eq.~(\ref{eq:lexp}) are then fixed by imposing this asymptotic power law, which yields
\be\label{alphas}
\frac{\alpha_n^{(1)}}{\alpha_n^{(0)}}=
\tR b_n\left[\frac{1}{b_n}\left(\frac{4+2n}{3}-a_n\right)-\frac{\alpha_{n-1}^{(1)}}{\alpha_{n}^{(0)}}\right]\,.
\ee
The first few coefficients reproduce the values of known transport coefficients \cite{Blaizot:2017lht,Teaney:2013gca}, for instance 
$\alpha_2^{(0)} = \frac{64}{105}\,\varepsilon\tau_R^2=\frac{4}{3}(\lambda_1+\eta\tau_\pi)$, $\alpha_1^{(1)} = -\frac{32}{315}\,\varepsilon\tau_R^2=\frac{4}{3}(\lambda_1-\eta\tau_\pi)$, with $\lambda_1$ and $\tau_\pi$ as defined in \cite{Baier:2007ix}.

\subparagraph{The attractor.}

\begin{figure}
\begin{center}
\includegraphics[width=0.3\textwidth] {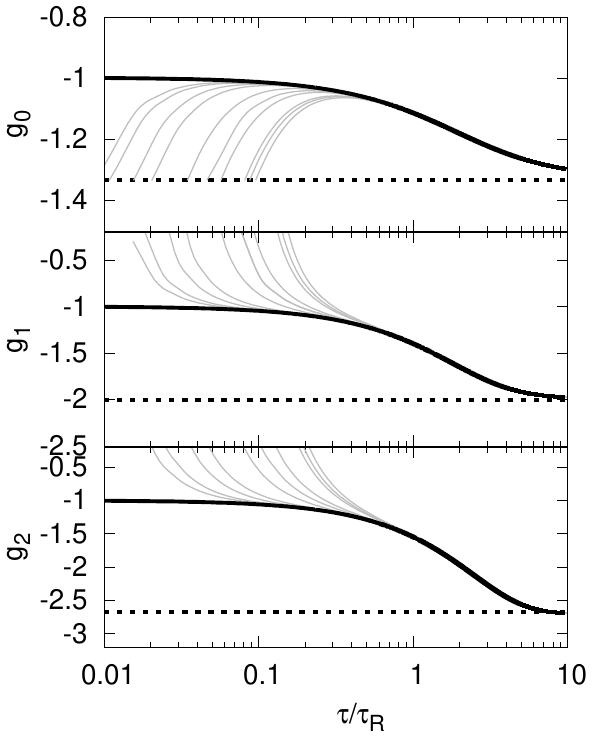}
\caption{Attractor solutions (black solid lines) to the ${\cal L}$-moment equations cut at $n<20$,
in terms of $g_0$, $g_1$ and $g_2$. Dotted lines correspond to the hydrodynamic
fixed point.
Solutions with random initial conditions are shown in grey.
\label{fig:att}
}
\end{center}
\end{figure}

One may define an attractor solution as the particular solution of Eqs.~(\ref{eq:eomL}) which, at short time, coincides with the free streaming fixed point $g_n=-1$, and at large time goes over to the hydrodynamic fixed point.  It can be determined numerically, by solving Eqs.~(\ref{eq:eomL}) with initial conditions specified by the constants (\ref{Ans}).  
 We have checked that $g_0$ obtained in this way is consistent with what was found by other methods in \Ref{Romatschke:2017vte,Heller:2015dha}.
The solution, obtained by truncating Eqs.~(\ref{eq:eomL}) at $n<20$, is displayed in Fig.~\ref{fig:att} for the first few $g_n(\tau)$. The universal character of the curves is worth emphasizing. 
All the $g_n$'s behave in the same way, interpolating between the two fixed point $g_n\approx -1$\footnote{Because of the truncation at $n<20$, the fixed point does not lie exactly at $-1$, but at $-1.00294$} and  $g_n=-(4+2n)/3$, the transition occurring when $\tau\sim\tau_R$. 

\subparagraph{Hydrodynamics.}

At this point, we note that the truncations of the equations (\ref{eq:eomL}) for the moments are closely related to successive viscous corrections to hydrodynamics. We have already seen that the lowest order truncation, i.e., with  only ${\cal L}_0$ non vanishing, is identical to ideal hydrodynamics. The truncation at order $n=1$ yields two coupled equations that can be cast in the form
\begin{align}\label{dnmr}
\partial_\tau\epsilon =-\frac{4}{3}\frac{ \epsilon}{\tau} + \frac{\Pi}{\tau}\,,\quad
\partial_\tau \Pi=\frac{4}{3}\frac{\eta}{\tau\tR}
-a_1\frac{\Pi}{\tau}-\frac{\Pi}{\tR}\,,
\end{align}
where 
$
\Pi\equiv -c_0{\cal L}_1$,
 and we used the leading order relation ${4}{\eta}/(3{\tau\tau_R)}=c_0b_1 {\varepsilon}/{\tau}.$
These are just the second order viscous hydrodynamic equations, in the version of Ref.~\cite{Denicol:2012cn} with $\beta_{\pi\pi}=a_1=38/21$.  The first order viscous hydrodynamics uses the solution of the second equation (\ref{dnmr}) for small $\tau_R$, viz. $\Pi\simeq 4\eta/(3\tau)=(16/45) \varepsilon (\tau_R/\tau)$. The much studied (lack of) convergence of the hydrodynamic gradient expansion in the context of Bjorken flow concerns the series of the coefficients  $\alpha_1^{(n)}$ in Eq.~(\ref{eq:lexp}) for ${\cal L}_1\sim \Pi$, as can be deduced from the solution of the coupled equations (\ref{dnmr}) at large time \cite{Spalinski:2017mel}.

Taking higher moments into account is tantamount to including higher order viscous corrections. For instance, the lowest order contribution of ${\cal L}_2$ to the equation for ${\cal L}_1$ reads
\be\label{hydrovisc3}
\frac{c_0 c_1 {\cal L}_2}{\tau}=\frac{c_1 b_2}{ c_0 b_1 \varepsilon} \frac{\Pi^2}{\tau}
\ee
where we have used Eqs.~(\ref{eq:lexp}) and (\ref{alphas}) to write ${\cal L}_2=\alpha_2{(0)}/\tau^2
=\alpha_2^{(0)}/(\alpha_1^{(0)} c_0)^2 \Pi^2$. It can be verified that the correction (\ref{hydrovisc3})  coincides with  the third order viscous correction derived in \Ref{Jaiswal:2013vta}. Obviously, it would be straightforward to obtain in this way higher order viscous corrections, if needed. Note that since $b_n\sim n$ at large $n$, $\alpha_n^{(0)}\propto n!$, and the series of the $\alpha_n^{(0)}$ suffers from the same lack of convergence as that of the $\alpha_1^{(n)}$ determining the viscous part of the energy momentum tensor.

\begin{figure}
\begin{center}
\includegraphics[width=0.30\textwidth] {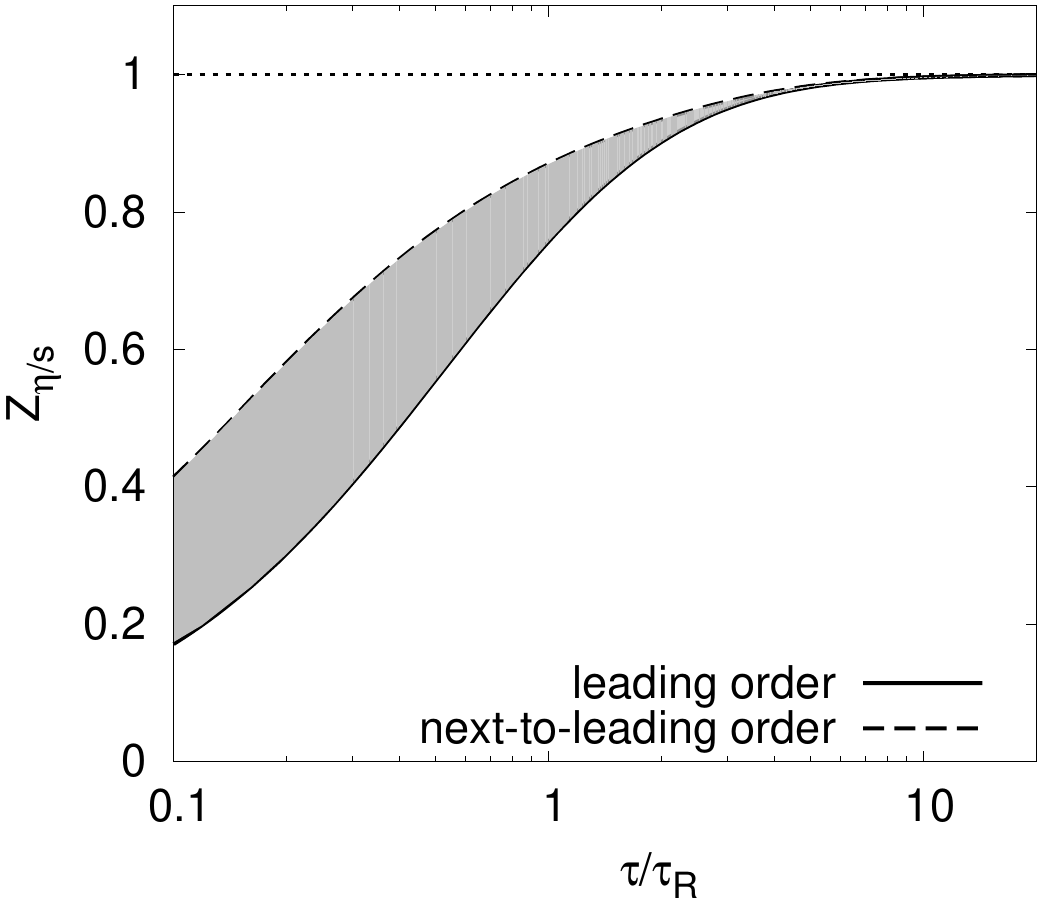}
\caption{ Renormalization constant $Z_{\eta/s}$ as a function of $\tau/\tau_R$. The leading order corresponds to Eq.~(\ref{L2L1}), the next-to-leading order include the correction due to $g_3(\tau)$. 
\label{fig:etasR}
}
\end{center}
\end{figure}

\subparagraph{Renormalization of $\eta/s$.}

Alternatively, the effects of the higher moments can be treated as a renormalization of the viscosity entering the equations for ${\cal L}_0$ and ${\cal L}_1$. To see that, rewrite the equation for ${\cal L}_1$ as
\begin{align}
\label{eq:etasR}
\partial_\tau {\cal L}_1 = -\frac{1}{\tau}\left(a_1{\cal L}_1 + b_1{\cal L}_0 \right)- \left[1+\frac{c_1 \tau_R}{\tau}
\frac{{\cal L}_2}{{\cal L}_1}\right]\frac{{\cal L}_1}{\tR},
\end{align}
with $Z_{\eta/s}^{-1}\equiv\left[1+\frac{c_1 \tau_R}{\tau}
\frac{{\cal L}_2}{{\cal L}_1}\right]$.
The dimensionless ratio ${\cal L}_2/{\cal L}_1$ is analytically related to the attractor $g_2(\tau)$,
the leading order result being
\be\label{L2L1}
\frac{{\cal L}_2}{{\cal L}_1}=-\frac{b_2}{a_2+{\tau}/{\tau_R}+g_2(\tau)}\,.
\ee
Sub-leading contributions involving higher $g_n$'s 
can be obtained iteratively.
The quantity $Z_{\eta/s}$  in  \Eq{eq:etasR} then defines a multiplicative renormalizaiton of $\eta/s$ (or equivalently of $\tau_R$: $\tau_R\to Z_{\eta/s} \tau_R$), whose variation with $\tau_R$ is displayed in 
\Fig{fig:etasR}. Since successive corrections alternate in sign, the grey band provides an estimate of the error. At large times, corresponding
to a system in local thermal equilibrium, $Z_{\eta/s}$ 
is close to unity. For systems far-from-equilibrium, $Z_{\eta/s}$ tends to
vanish. Thus, in systems out-of-equilibrium, 
higher order viscous corrections effectively  reduce the value of $\eta/s$ entering 
the second order viscous hydrodynamic equations, an effect first pointed out by  Lublinsky and Shuryak \cite{Lublinsky:2007mm}.
As can be seen on Fig.~\ref{fig:compL} (grey dashed line), this simple renormalization brings the solution of the lowest non trivial truncation quite close to the exact solution. That is, with this correction, second order viscous hydrodynamics reproduces accurately the exact solution of the kinetic theory. 

In summary, we have seen that it is possible for viscous hydrodynamics to describe accurately the evolution of  boost invariant plasmas, even in regimes where the usual conditions of applicability of hydrodynamics are not satisfied. This is because  the viscous hydrodynamic equations can be mapped into equations for moments of the momentum distribution that account exactly for the underlying kinetic theory.  Although the present discussion relies   on specific properties of Bjorken flow and the use of a simplified kinetic equation, we expect some general features to be robust, such as the existence of the free streaming and the hydrodynamic fixed points\footnote{In approaches based on holography, the hydrodynamic fixed point naturally emerges. However what plays the role of the free streaming fixed point in this context is unclear to us.}, joined by an attractor solution, or the  renormalization of the effective viscosity.  Clearly these results may have impact on the interpretation of heavy ion data and deserve further study. 

\section*{Acknowledgements}
LY is supported in part by the Natural Sciences and Engineering
Research Council of Canada. We thank Jean-Yves Ollitrault for insightful comments on the manuscript. 


\begin{thebibliography}{}

\bibitem{Florkowski:2017olj}
  W.~Florkowski, M.~P.~Heller and M.~Spalinski,
  arXiv:1707.02282 [hep-ph].

\bibitem{Heller:2011ju}
  M.~P.~Heller, R.~A.~Janik and P.~Witaszczyk,
  Phys.\ Rev.\ Lett.\  {\bf 108} (2012) 201602
  [arXiv:1103.3452 [hep-th]].


\bibitem{Romatschke:2017vte}
  P.~Romatschke,
  arXiv:1704.08699 [hep-th].
    
\bibitem{Heller:2015dha}
  M.~P.~Heller and M.~Spalinski,
  Phys.\ Rev.\ Lett.\  {\bf 115} (2015) no.7,  072501
  [arXiv:1503.07514 [hep-th]].
  
\bibitem{Spalinski:2017mel}
  M.~Spalinski,
  arXiv:1708.01921 [hep-th].

\bibitem{Strickland:2017kux} 
  M.~Strickland, J.~Noronha and G.~Denicol,
  arXiv:1709.06644 [nucl-th].
  
\bibitem{Romatschke:2017acs}
  P.~Romatschke,
  arXiv:1710.03234 [hep-th].

\bibitem{Denicol:2017lxn}
  G.~S.~Denicol and J.~Noronha,
  arXiv:1711.01657 [nucl-th].

\bibitem{Behtash:2017wqg}
  A.~Behtash, C.~N.~Cruz-Camacho and M.~Martinez,
  arXiv:1711.01745 [hep-th].
  
\bibitem{Bjorken:1982qr}
  J.~D.~Bjorken,
  Phys.\ Rev.\ D {\bf 27} (1983) 140.

\bibitem{Baym:1984np}
  G.~Baym,
  Phys.\ Lett.\  {\bf 138B} (1984) 18.
  
  
\bibitem{Blaizot:2017lht} 
  J.~P.~Blaizot and L.~Yan,
  JHEP {\bf 1711}, 161 (2017)
  doi:10.1007/JHEP11(2017)161
  [arXiv:1703.10694 [nucl-th]].

\bibitem{Bazow:2016oky}
  D.~Bazow, G.~S.~Denicol, U.~Heinz, M.~Martinez and J.~Noronha,
  Phys.\ Rev.\ D {\bf 94} (2016) no.12,  125006
  [arXiv:1607.05245 [hep-ph]].

\bibitem{Baier:2007ix}
  R.~Baier, P.~Romatschke, D.~T.~Son, A.~O.~Starinets and M.~A.~Stephanov,
  JHEP {\bf 0804} (2008) 100
  [arXiv:0712.2451 [hep-th]].


\bibitem{Teaney:2013gca}
  D.~Teaney and L.~Yan,
  Phys.\ Rev.\ C {\bf 89} (2014) no.1,  014901
  [arXiv:1304.3753 [nucl-th]].

\bibitem{Denicol:2012cn}
  G.~S.~Denicol, H.~Niemi, E.~Molnar and D.~H.~Rischke,
  Phys.\ Rev.\ D {\bf 85} (2012) 114047
   Erratum: [Phys.\ Rev.\ D {\bf 91} (2015) no.3,  039902]
  [arXiv:1202.4551 [nucl-th]].

  
\bibitem{Jaiswal:2013vta} 
  A.~Jaiswal,
  Phys.\ Rev.\ C {\bf 88}, 021903 (2013)
  [arXiv:1305.3480 [nucl-th]].


\bibitem{Lublinsky:2007mm}
  M.~Lublinsky and E.~Shuryak,
  Phys.\ Rev.\ C {\bf 76} (2007) 021901
  [arXiv:0704.1647 [hep-ph]].



\end{thebibliography}
\end{document}